# TRANSLATING THE CONCEPT OF GOAL SETTING INTO PRACTICE - WHAT 'ELSE' DOES IT REQUIRE THAN A GOAL SETTING TOOL?

Gábor Kismihók[1], Catherine Zhao[2], Michaéla C. Schippers[3], Stefan T. Mol[4], Scott Harrison[5], and Shady Shehata[6]

[1]*Leibniz Information Centre for Science and Technology, Hannover, Germany*
[2]*The University of Sydney, Sydney, Australia*
[3]*Erasmus University of Rotterdam, Rotterdam, Netherlands*
[4]*University of Amsterdam, Amsterdam, Netherlands*
[5]*Leibniz Institute for Research and Information in Education, Frankfurt, Germany*
[6]*YOURIKA, Waterloo, Canada*
Gabor.Kismihok@tib.eu, catherine.zhao@sydney.edu.au, mschippers@rsm.nl, s.t.mol@uva.nl,
Harrison@dipf.de, sshehata@yourika.ai

Keywords: Goal setting, self-regulated learning, learning intervention, curriculum, MOOC, higher education

Abstract: This conceptual paper reviews the current status of goal setting in the area of technology enhanced learning and education. Besides a brief literature review, three current projects on goal setting are discussed. The paper shows that the main barriers for goal setting applications in education are not related to the technology, the available data or analytical methods, but rather the human factor. The most important bottlenecks are the lack of students' goal setting skills and abilities, and the current curriculum design, which, especially in the observed higher education institutions, provides little support for goal setting interventions.

---

[1] https://orcid.org/0000-0003-3758-5455
[2] https://orcid.org/0000-0002-2791-4019
[3] https://orcid.org/0000-0002-0795-5454
[4] https://orcid.org/0000-0002-9375-3516
[5] https://orcid.org/0000-0001-6712-7784
[6] https://orcid.org/0000-0002-3258-6734

# 1 INTRODUCTION

Educational technology and 'big' data are having a major impact on learning these days: disruptive forces are modifying the modalities and strategies we choose to learn. Subsequently, the mastery of skills and competences enabling lifelong learning in the vast majority of aspects and fields of education are critically important in the 21st century (Ramsden, 2003, EUR-Lex, 2017). This movement is also visible in the area of Self-Regulated Learning (SRL), which has never been so actual and timely as it is these days (Archer, 1988; Schunk and Zimmerman, 2012). As a result, the needs of individual learners, and the integration of these needs into particular social and technical contexts play a more and more important role in contemporary education (Ferguson, 2012; Buckingham Shum and Ferguson, 2012).

Goal Setting (GS), as a critical and instrumental component of SRL (Pintrich, 2000), is suggested to be an important activity in learning intervention designs (Wise et al, 2014). Nevertheless, GS is still rarely used, especially in higher education despite its demonstrated positive effects on study success (Mol et al, 2016). Research also has shown already that through dashboards learners can visualise and internalize learning objectives (Scheffel et al, 2014; Verbert et al, 2014).

This paper sets out to rekindle discussions around GS to ensure that this important aspect of SRL gets attention and lands on the agenda of Technology Enhanced Learning (TEL) research and practice communities. To facilitate this conversation, we aim to summarize lessons learned from three recent European investigations in order to illustrate not only the potential, but also the pitfalls of GS. We also consider what should be the next steps for TEL researchers and practitioners to realize the power of GS. We hope that this paper will ignite dialogues within the TEL community about this important SRL concept, and that this will yield more studies and experiments in the near future.

To achieve this, this paper starts with a brief literature review on the current state of the art of GS. Then we discuss three GS investigations and their outputs, followed by a discussion on the bottlenecks and barriers facing GS research. We close our paper with suggesting future directions for GS stakeholders.

# 2 GOAL SETTING IN EDUCATION

## 2.1 State of the art

The principles of GS, which were developed in the mid-1960s by Edwin Locke, provide practical accounts of motivation in both managerial and academic contexts (Locke and Latham, 2006). Locke and his colleagues also showed that specific objectives lead to greater performance improvement than general ones. Furthermore, a linear relationship between goal difficulty and task performance has been established (Locke and Latham, 2006). Thus, GS is economical in financial terms, and has the potential to optimize task and academic performance (Schippers, 2017; Schmidt, 2013).

Recent studies also confirmed the importance of GS. Learning goals contribute to high interest, (Valle et al, 2017) and predict improvements in academic performance both in high school and higher education environments (Neroni et al, 2018, Burns et al, 2018; Schippers et al., 2020). In the area of educational computer games, GS increases comprehension (Erhel et al, 2019), especially when negotiation between learners is also facilitated (besides individual GS) (Chen et al, 2019), and affects the success of learners on the leaderboard (Landers et al, 2017).

In a more specific frame, GS can be an integral part of the feedback process that supports individual learning. For students in higher education, providing well defined feedback processes can enhance the learning process, especially when a formal GS protocol is included in the feedback cycle (Evans, 2013, Duffy and Azevedo, 2015). This also needs to be done tactfully, without affecting the student's ego: "Self-efficacy influences motivation and cognition because it affects students' task interest, task persistence, the goals they set, the choices they make and their use of cognitive, meta-cognitive and self-regulatory strategies" (Van Dinther et al, 2010, p. 97).

This reinforces the importance of understanding the student's state of mind and willingness to undertake GS as a learning strategy (Lazowski and Hulleman, 2016). "When students believed that they could get smarter over time, they were more likely to believe that working hard could help them succeed in school and they endorsed the goal of learning from coursework. These beliefs and goals motivated greater use of effective learning strategies" (Yeager and Walton, 2011, p. 286). Since GS can hence be seen as an effective strategy for improving learning trajectories, the question arises: what are the major obstacles to the more widespread adoption of GS in higher education?

We have seen that scepticism of psychological intervention studies is prudent where potential bias can be introduced, either through limited sample sizes or where incentives artificially inflate engagement. For example, Chase et. al (2013) constructed an experiment testing the effects of GS under the condition of values training. Students recruited for the experiment (N=132) had the opportunity to "win" goods with tangible value. Importantly, this study found that GS alone had no effect on learning trajectories. Only when values training was included did students perform better, thus putting scepticism in the centre of the issues of interventions' scalability and innovations, which facilitate GS.

In sum, there is evidence to show that GS can improve students' learning trajectories and outcomes. However this evidence needs to be critically challenged to best understand, what dimensions of the GS process can be scaled, to provide support above and beyond small scale interventions (Schippers & Ziegler, 2019).

## 2.2 Three cases of goal setting experimentations and deployment

With the support of educational technology, designing and running GS interventions – also on large (institutional) scale – is possible. To demonstrate this, in this paper we examine three recent attempts, which use GS in two different educational settings (higher education and MOOC) to investigate the relationship between GS and learning outcomes. As it was elaborated in these studies, researchers face a range of problems, when it comes to motivating learners to set and to monitor their goals throughout their learning journey. Therefore, we aim to shed light on the criticality of the educational context, with a focus on how decisions have led to different outcomes in these studies (see table 1).

Schippers and her team (Schippers et al, 2015; 2020) designed a three staged GS intervention (with a GS application) that scaffolds the GS process for university first year students (n=2928) and encourages them to achieve their goals. The intervention requires students to start explicitly to conceptualize by writing their desired future (in stage 1), and to articulate a step-by-step plan for achieving their goals (in stage 2). Alongside these procedures, students are encouraged to assess practicality and attainability of their goals, in order to stay on the 'right' track. At the operational level the intervention is an integral part of the curriculum across campus, despite the fact that it is technically a stand-alone system. The studies by Schippers et al. (2020) show that participation in the intervention closed the gender and ethnicity achievement gap (Schippers et al, 2015). Further, the results indicate that formal participation (e.g. an element of the assessment task) in the intervention, the amount of writing and the quality of the writing are the three key factors that determine the effectiveness of GS, whilst whether or not students set academic goals does not seem to matter. In other words, the process by which students engage psychologically in setting goals makes the difference – in that it enhances the student's self-efficacy, optimizes effort, and psychologically better prepares them to achieve their goals (Schippers, 2017; Schippers et al, 2017).

Table 1. Comparison of three goal-setting studies

| Element | Dimension | Schippers et. al | ProSOLO | Mol et. Al |
|---|---|---|---|---|
| **Intervention setup** | *Educational context* | University program based | MOOCs | University program based |
| | *Learner participation* | Opt-in informed consent | Optional | Opt-in informed consent |
| | *Dashboard* | No | Yes | Yes |
| **Learners' prior experiences** | *Background of targeted learners* | University students | Corporate professionals | University students |
| | *Assumed learner skills* | Metacognitive skills | Not specified | Not specified |
| **Goal related activities** | *Engagement time* | Stage 1&2, 10 min, photography in stage 3 | Not specified | Not specified |
| | *Means to set goals* | Write own goals | Write/Adopt external pre-specified goals | Write/Adopt peer's goal |
| | *Criteria for setting goals* | Practical & attainable | No | SMART |
| **Feedback** | *Instructor feedback to students* | No | No | Ratings of goals against criteria |
| **Support** | *Peer-student support* | No | Yes | Depending on student's choice |
| | *Other support* | Scaffolding through 'steps' in the system | Ad-hoc inquiry & technical support | Ad-hoc inquiry & technical support |
| **Outcome** | *Anticipated outcome* | A package intervention: "life crafting" | Foster effective self-regulated learning | Improved academic performance |
| | *Actual outcomes* | Enhanced student well-being and performance | Learners' uncertainty | Low participation |

The GS intervention designed by Mol and colleagues (Kobayashi et al, 2017) investigates the simultaneous effect of GS on university students' approaches to learning, and their academic performance. The study adopts the SMART (Specific, Measurable, Attainable, Realistic and Time-bound) characteristics as criteria (Conzemius and O'Neill, 2009; O'Neill, 2000) that guide students in setting effective goals. They also developed a GS tool, in which students could compose their goals, and append these with deadlines. This GS tool was connected to a Learning Record Store (LRS) which also set up to record additional data from the Learning Management Systems (LMS) about students' actual performance during the course. The study involved one university course at the University of Amsterdam and courses at three Australian-based universities. In the first lecture, students were introduced to GS theory with an emphasis on its benefits, and the custom developed tool and its key features. Specifically, the tool 1) allows students to set multiple main goals and associated sub-goals, (based on the fact that one oftentimes pursues multiple goals simultaneously (Austin, and Vancouver, 1996)); 2) students can view and adopt each other's goals, but only those that are made 'public' by the students who set them; 3) students can edit or delete goals after setting them; 4) students can view properties of goals e.g. structure, deadlines through a dashboard, against the timeline of their university course; 5) instructors can rate the quality of goals against the SMART criteria, students can view the ratings should the goal be made public. The intervention however has attracted low student participation in the pilot stage. The authors speculate that reasons of, 1) GS being optional, and as such independent of the course curriculum and not rewarded with course credit, 2) variability in instructors and tutors understanding of GS, 3) lack of student support (e.g. feedback) and GS learning resources, may have contributed to this outcome. Furthermore, the informed consent procedure that was employed, may have unintentionally scared some students off, as it also requested access to their LMS data and assessment outcomes. This latter issue also ties into the larger question of whether GS interventions should be positioned as a teaching tool, a research project, or both. Framing GS in terms of a teaching resource, may enhance face validity in the eyes of students, although evidencing such interventions is clearly more of a research question.

Gasevic and colleagues (Rosé et al, 2015; Jo et al, 2016; Jo et al, 2016) implemented the ProSOLO system to encourage learners to set goals and to foster social learning in a Massive Open Online Learning Course (MOOC) called Data, Analytics, and Learning. It targets corporate working professionals, who are assumed to have a reasonably high level of digital literacy. Compared to university campus-based courses, MOOCs generally target educated adult learners from much more diverse demographic backgrounds and with a wider range of motivations. Furthermore, there is evidence that GS predicts the attainment of course objectives (Kizilcec et al, 2018). Thus in the highly autonomous learning space, ProSOLO is designed to personalize the development of competencies, which are mapped to learning activities throughout the MOOC. Learners are encouraged to set up their own space that is comprised of predefined competencies (by course instructors) they want to develop, or their own learning goals (if the competencies do not match), a social network they can build by being able to follow one another through social media, and a learning progress feature. Learners are expected to link this personalized hub to the assignment submission process on the MOOC platform, toward course completion. This approach provides opportunities for learners who intend to purchase a certificate to demonstrate competencies with 'evidence'. However, the patterns of MOOC learners' engagement with the system, and their discussions in the MOOC forum point out a number of problems: 1) some learners seem to be confused with regard to having to engage with both the MOOC platform and ProSOLO; 2) some learners were not familiar with the technology; and more importantly, 3) despite high autonomy in MOOCs, when learners are not able to make informed choices of how to effectively learn, they fall back to what they are familiar with, which is oftentimes, a linear learning progression and a structured instructional norm, rather than the social construction of knowledge.

## 3 DISCUSSION

### 3.1 Lessons learned from goal setting experiments

This paper unpacked the TEL related GS literature and reviewed three technology-enhanced interventions to bring forward the dimensions that are believed to be important to the success of GS interventions in education. From this analysis, it emerged that the course instructor, peer learners, and the goal-setting interface designer play key roles in shaping the learner's perception of GS from the outset. In the studies where researchers were course instructors, the GS concept was 'translated' more effectively into meaningful actions and thinking processes that were relevant to students. However, it is worth considering the relationship between GS and assessment. In both Schipper's intervention and the ProSOLO experiment, GS is an element of assessment (despite having a minimal or no grade attached). The possible explanation of the difference is that assessment matters in university learning but not in a MOOC. Furthermore, the two interventions are very different: The one used by Schippers and colleagues is based on expressive writing and personal GS, (also referred to as "life crafting", Schippers & Ziegler, 2019), while ProSolo is aimed at competency development. Future research should investigate this relationship carefully.

Less apparent is the broader context in which more distal stakeholders come into the play. These include the nature of the learning episode (e.g. a course, a program, or a MOOC), the technological readiness of the offering university, and the institutional culture. Meanwhile, the forms of education are becoming more diverse, which attract learners with diverse motivations to learn. Especially in MOOCs, learning is often not tied to assessment (Jordan, 2015; Vigentini and Zhao, 2016). While what drives learning in MOOCs is debatable, future GS research should respond to the challenge of how to integrate GS into a personalized learning journey with the support of analytics.

How to implement a technology-enhanced GS intervention for students to make learning more effective does not have a straight answer. While the design and delivery process is complex, the debate remains an educational one - how does the student benefit from setting goals in a university course, or a degree program? Furthermore, how may researchers effectively demonstrate the value of GS to university stakeholders, to initiate system development and (re-)configuration that lays the technical foundation for TEL to empower GS interventions?

On the other hand, the program conveyor's perception of what matters most to developing graduate capabilities may determine the scale at which a GS intervention is implemented (e.g. in an individual course vs. core courses) throughout the program. Secondly, institutional technology readiness directly impacts on how a GS tool can be integrated into other university supported systems. Thus university culture to an extent shapes the way the student learns and what the teacher teaches. To this end, researchers should consider, where GS fits in an educational experience that is unique to the institution. Figure 1 presents these influences at different levels on GS interventions.

Figure 1. The multi-layer framework for an effective technology-enhanced goal-setting intervention

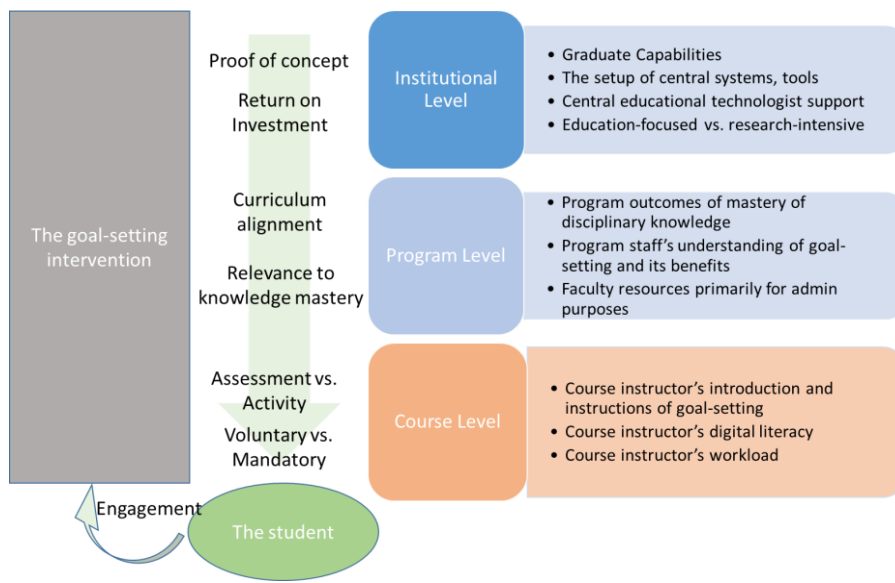

## 3.2 Bottlenecks for goal setting in higher education

GS in general is "a short and seemingly simple intervention (that) can have profound effects" (Wilson, 2011), and it has been supported a number of times in the past (Morisano et al, 2010, Travers et al, 2015, Schippers et al, 2020). However, there are several reasons why GS implementations in higher education can fail. Here we will focus on discussing the three most important potential bottlenecks.

The first bottleneck is the lack of ability from the student side to self-regulate and set goals. This has been confirmed by a previous study (McCardle et al, 2017), and it has been especially apparent in the Mol et al. pilot, where students failed to come up with goals altogether. Here students' looked at GS as an extra assignment on top of their curricular work, which does not help their progress, but only limits the time to spend on reaching the course objectives. Researchers think that this is a critical point. It is very difficult to direct students towards setting their own goals in relation to a course or a learning programme, if those goals are already set by the organization or the teacher. What happens in this case is, that students simply copy those course objectives and spend very little time about thinking and operationalizing their own self developed objectives – which would be the real benefit of GS. When this happens, goals are set to be unrealistic and they fail to consider resources or capabilities.

Furthermore, when students actually set goals, oftentimes they lack the ability to evaluate crucial information about the obstacles and challenges that they face, in achieving their goals. Despite the evidence that SRL supports cognitive and meta-cognitive abilities of students (Thomas, et al, 2016), in a learning environment, where students are pushed into a reactive rather than a proactive role when it comes to designing and controlling their own learning, GS can play only a marginal role. To overcome this problem, in the intervention used by Schippers, students had to come up with a detailed plan to overcome obstacles and challenges.

The second bottleneck is a more methodological one. From the available literature and experiments it is not obvious, what the best methods are to incorporate GS in course design (in various contexts). As it was mentioned earlier, it is very difficult to implement effective GS mechanisms in the curricula, if learning goals are already pre-developed and made available for the course participants beforehand. Methods need to be in place to co-develop these course objectives with the students, which require more flexible curricula. Nevertheless, the design of GS interventions may share some similarities with other educational approaches such as the use of e-Portfolios (Berg et al, 2018) to develop a 'learning journey'.

The third methodological issue is about rewarding students, who actually set goals. According to the pilots, oftentimes students do not believe that the rewards they will receive for goal accomplishment are worth the effort that they need to invest to achieve them. For instance, when there are too many goals to achieve, a mechanism should be in place to prioritize certain goals over others. In the case of the successful Schippers' intervention, it was shown that setting personal goals has a rewarding effect on students. However, the skill of setting (personal) goals effectively is not an easy one to master, and training this skill should not only happen in higher education, but also much earlier in primary and secondary education.

Thus the authors suggest further opportunities for teaching academics to gain a more thorough understanding of the concept and practice of GS through professional development programs. This skill is not only important for students, but also for their teachers and indeed researchers.

## 3.3 Integrating goal setting in the academic program

Given that GS enhances study success, the next question is how to make sure that as many students profit from this intervention as possible (Schippers, 2017). However, if the GS intervention is made optional, students may not engage with it, especially the poor performers who may stand to benefit most. It was learned from one of the abovementioned pilots that when the third part of the intervention was made optional, from 1,200 students, only 45 students participated in that third part! Therefore, it may be important to make the GS intervention part of the curriculum, so both students and educational institutions benefit from the positive outcomes (Schippers, 2020; Clonan et al, 2004). GS may be notably useful when learners are in a transitional period of their lives, as for instance progressing from school to higher education (Schippers & Ziegler 2019; Wilson, 2011), or from higher education to the labour market (Schippers, 2020; Berg et al, 2018). However the effects of making GS mandatory should be further investigated, as ownership is critical to the success of GS.

A positive outcome from the pilots is that technical infrastructure, for collecting and analysing learning related data in relation to goals is, in general, not perceived as a bottleneck in GS experimentations.

## 4 CONCLUSIONS

GS has a number of advantages, when it comes to applications in a number of educational contexts. The method is easy to implement from a technical point of view, and it works well together with existing educational and analytical technologies. Evidence also shows that GS can significantly improve both the self-regulation, and the academic performance of learners. However there are a number of barriers on the human side, which still need substantial efforts to overcome. The most important barriers are the low levels of student abilities to set goals, and the current – especially in traditional classroom settings – methods for pre-defining learning outcomes for learners and classes. It comes without saying that these issues need further investigation.

The authors think that teaching and research communities should engage in more in depth conversations about GS in order to understand and use this concept better in the future. Therefore, the most important aim of this paper was to provide ammunition for these discussions by highlighting the above mentioned critical observations. On a positive note, the authors of this paper strongly believe that, especially in the light of the ongoing GS experiments and implementations, there is a bright future for GS in education.

## REFERENCES

Ames, C., Archer, J. 1988. Achievement goals in the classroom: Students' learning strategies and motivation processes. *Journal of Educational Psychology* 80, 3 (1988), 260–267.

Austin, J. T., & Vancouver, J. B. 1996. Goal constructs in psychology: Structure, process, and content. Psychological bulletin, 120(3), 338.

Berg, A.M., Branka, J., Kismihók, G. 2018 Combining Learning Analytics with Job Market Intelligence to Support Learning at the Workplace. *In: Digital Workplace Learning.* pp. 129–148. Springer, Cham. https://doi.org/10.1007/978-3-319-46215-8_8.

Burns, E.C., Martin, A.J., Collie, R.J. 2018. Adaptability, personal best (PB) goals setting, and gains in students' academic outcomes: A longitudinal examination from a social cognitive perspective. *Contemporary Educational Psychology.* 53, 57–72

Chase, J. A., Houmanfar, R., Hayes, S. C., Ward, T. A., Vilardaga, J. P., Follette, V., 2013. Values are not just goals: Online ACT-based values training adds to goal setting in improving undergraduate college student performance, *In Journal of Contextual Behavioral Science*, Vol 2, Iss 3–4, pg. 79-84

Chen, Z.-H., Lu, H.-D., Chou, C.-Y. 2019. Using game-based negotiation mechanism to enhance students' goal setting and regulation. *Computers & Education.* 129, 71–81

Clonan, S. M., Chafouleas, S. M, McDougal, J. L., Riley-Tillman T. C. 2004. Positive psychology goes to school: Are we there yet? Psychology in the Schools 41, 1, 101–110.

Conzemius A., O'Neill, J., 2009. *The Power of SMART Goals: Using Goals to Improve Student Learning.* Solution Tree Press.

Duffy, M.C., Azevedo, R. 2015. Motivation matters: Interactions between achievement goals and agent scaffolding for self-regulated learning within an intelligent tutoring system. Computers in Human Behavior. 52, 338–348.

Erhel, S., Jamet, E., 2019. Improving instructions in educational computer games: Exploring the relations between goal specificity, flow experience and learning outcomes. Computers in Human Behavior. 91, 106–114.

EUR-Lex - c11090 - EN - EUR-Lex. 2017. http://eur-lex.europa.eu/legal-content/EN/TXT/?uri=LEGISSUM:c11090

Evans, C., 2013 Making Sense of Assessment Feedback in Higher Education, In Review of Educational Research, Vol. 83, No. 1, pp. 70–120,

Ferguson, R.,2012. Learning analytics: drivers, developments and challenges. *International Journal of Technology Enhanced Learning* 4, 5–6, 304–317.

Jo, Y., Tomar, G., Ferschke, O., Rosé, C. P., Gasevic, D. 2016. Expediting support for social learning with behavior modeling. arXiv preprint arXiv:1605.02836.

Jo, Y., Tomar, G., Ferschke, O., Rosé, C. P., & Gašević, D., 2016. Pipeline for expediting learning analytics and student support from data in social learning. *In Proceedings of the Sixth International Conference on Learning Analytics & Knowledge*, 542-543. ACM.

Jordan. K., 2015. Massive open online course completion rates revisited: Assessment, length and attrition. *The International Review of Research in Open and Distributed Learning* 16, 3 (2015).

Kizilcec, R.F., Pérez-Sanagustín, M., Maldonado, J.J., 2017 Self-regulated learning strategies predict learner behavior and goal attainment in Massive Open Online Courses. *Computers & Education.* 104, 18–33.

Kobayashi, V., Sanagavarapu, P., Zhao, C., Mol, S.T., & Kismihok, G., 2017. Investigating the relationships among self-regulated learning, approach to learning, goal orientation, LMS activity and academic performance. In *Proceedings of the 1st Learning & Student Analytics Conference: Implementation, Institutional Barriers and New Development*, Amsterdam, the Netherlands.

Landers, R.N., Bauer, K.N., Callan, R.C., 2017 Gamification of task performance with leader-boards: A goal setting experiment. *Computers in Human Behavior.* 71, 508–515.

Lazowski, R. A., Hulleman, C. S. 2016 Motivation Interventions in Education: A Meta-Analytic Review, *In Review of Educational Research,* Vol. 86, No. 2, pp. 602– 640

Locke E. A., Latham, G. P., 2006. New directions in goal-setting theory. *Current directions in psychological science* 15, 5, 265–268.

McCardle, L., Webster, E.A., Haffey, A., Hadwin, A.F 2017. Examining students' self-set goals for self-regulated learning: Goal properties and patterns. *Studies in Higher Education.* 42, 2153–2169.

Mol, S.T., Kobayashi, V.B., Kismihók, G. and Zhao, C. 2016. Learning through goal setting. *In Proceedings of the Sixth International Conference on Learning Analytics & Knowledge,* 512–513

Morisano, D., Hirsh, J. B., Peterson, J. B., Pihl, R. O., & Shore, B. M. 2010. Setting, elaborating, and reflecting on personal goals improves academic performance. *Journal of Applied Psychology*, 95(2), 255–264.

Neroni, J., Meijs, C., Leontjevas, R., Kirschner, P.A., De Groot, R.H.M. 2018. Goal Orientation and Academic Performance in Adult Distance Education. *irrodl.* 19.

O'Neill, J., 2000. SMART Goals, SMART Schools. *Educational Leadership* 57, 5 (2000), 46–50.

Pintrich, P.. 2000. The role of goal orientation in self-regulated learning. *Handbook of self-regulation* 451, (2000), 451–502

Ramsden,P., 2003. *Learning to Teach in Higher Education.* Routledge

Rosé, C. P., Ferschke, O., Tomar, G., Yang, D., Howley, I., Aleven, V. & Baker, R. 2015. Challenges and opportunities of dual-layer MOOCs: Reflections from an edX deployment study. *In Proceedings of the 11th International Conference on Computer Sup-ported Collaborative Learning* (CSCL 2015) (Vol. 2).

Scheffel, M., Drachsler, H., Stoyanov, S., Specht, M. 2014. Quality in-dicators for learning analytics. *Journal of Educational Technology & Society* 17, 4, 117.

Schippers, M. C. 2017. *IKIGAI: Reflection on life goals optimizes performance and happiness* (EIA-2017-070-LIS ed.). Rotterdam: Erasmus Research Institute of Management

Schippers, M. C., Morisano, D., Locke, E. A., Scheepers, A. W. A., Latham, G. P., & de Jong, E. M. 2020. Writing about personal goals and plans regardless of goal type boosts academic performance. *Contemporary Educational Psychology*, 60, 101823.

Schippers, M. C., Scheepers, A., Morisano, D., Locke, E. A., & Peterson, J. B. 2017. Conscious goal reflection boosts academic performance regardless of goal domain. Manu-script submitted for publication.

Schippers, M. C., & Scheepers, A. W. & Peterson, J. B. 2015. *A scalable goal-setting intervention closes both the gender and minority achievement gap.* Palgrave Communications 1:15014

Schippers, M.C. & Ziegler, N. 2019. Life crafting as a way to find purpose and meaning in life. *Frontiers in Psychology*, 10(2778)

Schmidt. F. I. 2013. *The economic value of goal setting to employers.* New develop-ments in goal setting and task performance. 16–20.

Schunk, D. H. Zimmerman. B. J., 2012. *Motivation and Self-Regulated Learning: Theory, Research, and Applications.* Routledge.

Buckingham Shum, S., Ferguson, R. 2012. Social learning analytics. *Journal of educational technology & society* 15, 3, 3.

Thomas, L., Bennett, S., Lockyer, L. 2016. Using concept maps and goal-setting to support the development of self-regulated learning in a problem-based learning curriculum. *Medical Teacher.* 38, 930–935.

Travers, C. J., Morisano, D., & Locke, E. A. 2015. Self-reflection, growth goals, and academic outcomes: A qualitative study. *British Journal of Educational Psychology*, 85(2), 224–241.

Valle, A., Núñez, J.C., Cabanach, R.G., Rodríguez, S., Rosário, P., Inglés, C.J. 2015. Motiva-tional profiles as a combination of academic goals in higher education. *Educational Psychology.* 35, 634–650.

Van Dinther, M., Dochy, F., & Segers, M., 2010, Factors affecting students' self-efficacy in higher education, *In Educational Research Review,* Vol 6, Iss 2, Pages 95-108

Verbert, K., Govaerts, S., Duval E.,, Santos J. l.,, Van Assche, F., Parra, G. and Klerkx, J.. 2014. Learning dashboards: an overview and future research opportunities. *Personal and Ubiquitous Computing* 18, 6 (2014), 1499–1514.

Vigentini, L., Zhao, C., 2016. Evaluating the 'Student' Experience in MOOCs. *In Proceedings of the Third (2016) ACM Conference on Learning@ Scale*, 161–164.

Wilson, T. 2011. *Redirect: The surprising new science of psychological change*. Penguin UK.

Wise, A., Zhao, Y., Hausknech, S.T. 2014. Learning analytics for online discussions: Embedded and extracted approaches. *Journal of Learning Analytics* 1, 2, 48–71.

Yeager D. S., & Walton, G. M. 2011. Social-Psychological Interventions in Education: They're Not Magic. *In Review of Educational Research,* Vol. 81, No. 2, pp. 267-301.